# The V-Lab VR Educational Application Framework

XR2Learn V-Lab

A Beacon Application of the XR2Learn Project

Vasilis Zafeiropoulos

School of Science and Technology, Hellenic Open University, Patras, Greece, vasiliszaf@eap.gr

George Anastassakis

School of Science and Technology, Hellenic Open University, Patras, Greece, ganast@ganast.com

Theophanis Orphanoudakis

School of Science and Technology, Hellenic Open University, Patras, Greece, fanis@eap.gr

Dimitris Kalles

School of Science and Technology, Hellenic Open University, Patras, Greece, kalles@eap.gr

Anastasios Fanariotis

School of Science and Technology, Hellenic Open University, Patras, Greece, afanariotis@eap.gr

Vassilis Fotopoulos

School of Science and Technology, Hellenic Open University, Patras, Greece, vfotop1@eap.gr

*This paper presents the V-Lab, a VR application development framework for educational scenarios mainly involving scientific processes executed in laboratory environments such as chemistry and biology laboratories. This work is an extension of the Onlabs simulator which has been developed by the Hellenic Open University as a distance teaching enabler for similar subjects, helping to alleviate the need for access to the physical laboratory infrastructure; thus, shortening training periods of students in the laboratory and making their training during the periods of physical presence more productive and secure. The extensions of the Onlabs to deliver an enhanced and modular framework that can be extended to multiple educational scenarios is the work performed within the context of the European project XR2Learn (Leveraging the European XR industry technologies to empower immersive learning and training).*

## 1 INTRODUCTION

Science university departments are confronted with major difficulties regarding their students' laboratory training. Such difficulties are the sensitivity of the equipment and lab instruments, which restrict their use, and the risk of accidents, which also prevents them from being fully explored. The problem becomes greater in the case of distance education, such as offered by Open Universities, in which the students visit the lab facilities rarely and do not have the opportunity to make thorough use of the lab equipment. For these reasons, Hellenic Open University, offering distance education in science, has developed Onlabs, a 3D virtual laboratory for science undergraduate students to train themselves before they use the on-site lab. In Onlabs, students have the opportunity to make free use of the simulated instruments and conduct virtual experiments, enabling learning by trial-and-error. Since the beginning of 2023, a new version of our virtual lab has been being developed, namely V-Lab, for the purposes of the XR2Learn project[1]. V-Lab offers the same functionality with Onlabs but with enhanced, state-of-the-art 3D graphics and a new, user-friendlier interface, which aspires to make the training process even more effective. In addition, it encompasses a modular architecture which facilitates the process of developing and adapting more training scenarios.

## 2  BACKGROUND AND RELATED WORK

Interactive computer-based applications for science and biology learning has in the past been developed and tested among students and claimed encouraging learning results; 212 junior high school students (13-14 years old) in Greece were provided with an interactive 3D animation, accompanied by narration and text, dealing with "methods of separation of mixtures" which in general, did increase the students' interest in science [2], while 44 magnet science and medical technology high school students (17-18 years old) in Texas, USA, improved their molecular biology skills by using a computer-based simulation designed for training in the production of a transgenic mouse model, independently of their previous knowledge of it [5]. Moreover, a virtual world under the name of Multiplayer Educational Gaming Application (MEGA) was designed for and used by 131 US college prep students, not of the highest performance in their cohort, in which the latter had to solve a CSI-like murder case using their skills of science and scientific inquiry, i.e., understanding of pedigrees, Mendelian inheritance, blood types and DNA fingerprinting; eventually, 94% of the participant students practiced successfully their basic scientific skills to solve the case while all but three students positively finished the game within a 90-minute class [1].

Virtual labs, also known as online laboratories or remote laboratories, have emerged as a powerful tool in the field of education and research. These digital environments provide learners with hands-on experiences and experimental simulations that mirror real-world laboratory settings. By leveraging advancements in technology and connectivity, virtual labs offer numerous benefits such as increased accessibility, scalability, and cost-effectiveness. Such realistic and instructive virtual labs are Labster, developed by the Danish multi-national company of the same name[2], and Learnexx 3D, developed by Solvexx Solutions Ltd, based in the UK[3].

Much research has focused on the development and evaluation of virtual labs across various disciplines; for example, the effectiveness of virtual labs in teaching chemistry concepts to school and undergraduate students has been thoroughly examined. The learning outcomes of students who used virtual labs has been compared with those who engaged in traditional laboratory sessions. The results demonstrated that virtual labs indeed facilitated effective knowledge acquisition but also promoted deeper understanding and engagement among the learners [6].

Furthermore, virtual labs have also made significant contributions to engineering education. Recently, a virtual lab platform that allowed students to design, simulate, and test electronic circuits remotely was developed. Through a series of experiments, the effectiveness of the platform in improving students' practical skills and problem-solving abilities was evaluated. The results indicated that virtual labs not only provided students with a safe and controlled learning environment but also enabled them to gain valuable hands-on experience, comparable to traditional in-person laboratories [3].

---

[1] Extended Reality for Learning in New Dimensions: https://xr2learn.eu.
[2] https://www.labster.com/
[3] http://learnexx.com/

Overall, these recent works highlight the growing interest in virtual labs and their potential to revolutionize education and research. By leveraging technologies such as simulation and virtual reality, virtual labs offer a wide range of benefits, including enhanced learning outcomes, increased accessibility, and improved engagement.

## 3 OVERVIEW OF ONLABS

### 3.1 Development

Onlabs has been being developed as a 3D simulation using two high-end 3D game-engines; from 2012 to 2015 using Hive3D, a game engine released by Eyelead Ltd, a computer game company in Athens, Greece[4]; and from 2016 until today using the Unity engine. While versions under Hive3D were developed for Windows, versions under Unity run on other operating systems as well.

Apart from the afore-mentioned student versions, in October 2020, we also released Onlabs Machine Learning version 1.0, which, as its name suggests, is used for machine learning training with respect to the software's embedded assessment mechanism. The latest versions of Onlabs can be download from our website[5]. More info about the earlier stages of Onlabs development are given in previous works of ours [7][8].

Figures 1 and 2 show respectively a view of HOU's on-site biology lab and a screenshot of its virtual simulation, Onlabs.

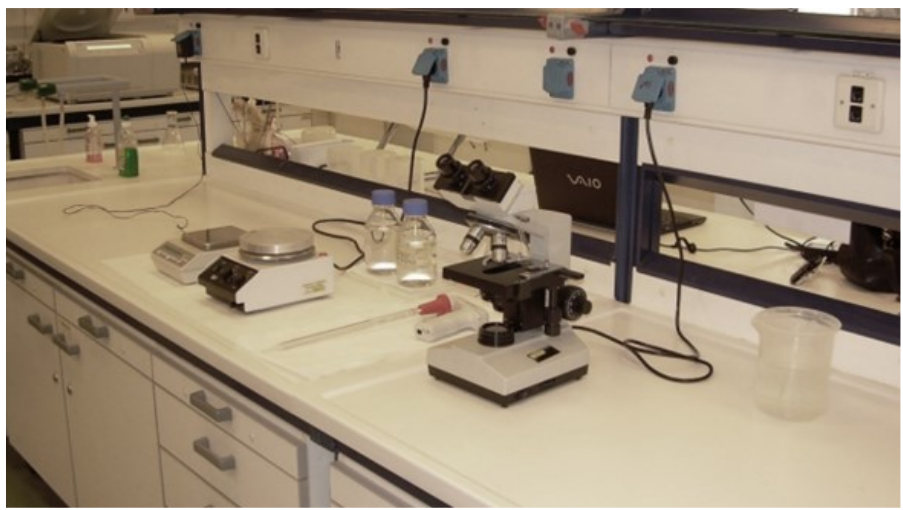

Figure 1. On-site biology laboratory of Hellenic Open University

---

[4] https://www.eyelead.com/
[5] http://onlabs.eap.gr/

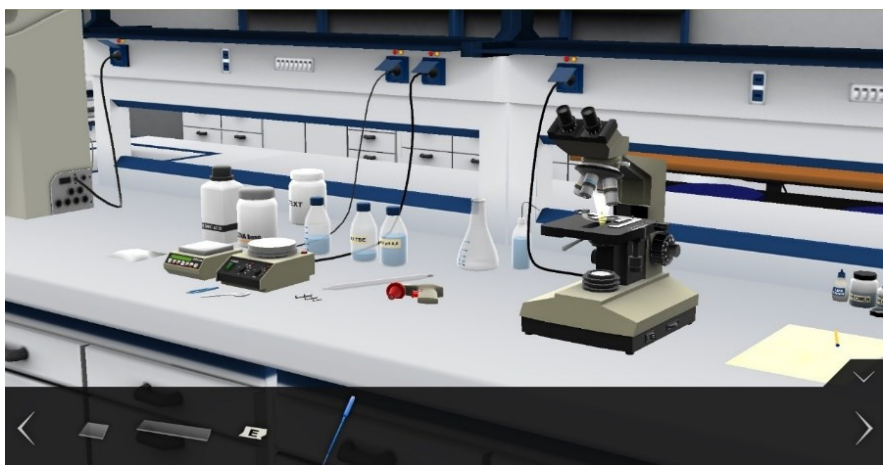

Figure 2. Virtual laboratory of Hellenic Open University (Onlabs)

**3.2 Playing Modes**

Onlabs's latest stable version, v2.1.2, contains two distinct simulated experimental procedures and three different modes of playing. The procedures are those of *microscoping of a test specimen*, in which the user is supposed to set up the microscope, create a test specimen and microscope it with all objective lenses of the photonic microscope; and the *preparation of 500ml of 10X TBE water solution*, for which the user weights 17.4gr of Boric Acid and 54gr of Trizma Base powders on the electronic scale, dissolves them in water with the magnetic stirrer and adds extra water as well as EDTA pH 8.0 to the produced solution. The playing modes are those of *Instruction*, where the user is instructed by voice and text in completing the selected procedure, each time being allowed to perform only the suggested move; *Evaluation*, where the user can make any move they want within the selected procedure and receives an evaluation for their performance; and *Experimentation*, where the user may make any action they want (provided it is included in the game) and handles all the instruments and equipment from both procedures without being evaluated.

Onlabs version 3.0 contains two new experimental procedures which respectively supersede the procedures in version 2.1.2; *microscoping*, like in the previous versions, and *electrophoresis*, whose first part is the preparation of 500ml of 10X TBE water solution. But while microscoping is available in both modes of Instruction and Evaluation, electrophoresis is in neither of them; however, electrophoresis instruments with their full functionality are included in the Experimentation Mode along with those of microscoping.

**3.3 Interface**

In Onlabs, the interaction between the user and the environment is similar to the one of adventure games; the user makes use of the arrow keys to navigate in it and the mouse to press buttons, turn knobs and use specific objects alone or in pairs. Furthermore, they are equipped with an inventory in which they can collect collectable objects. The actions permitted on the various objects are *press* (valid for switches), *rotate* (valid for knobs), *zoom* (valid for instruments to lay focus on), *pull* (valid from plugs), *pick up* in the inventory and *joint use* with another object.

In V-Lab however, the inventory has been abolished for the purpose of realism; while the inventory is useful as an abstract place for objects to kept while navigating around, it is not realistic at all, since the size of several collectable objects and instruments is too big. Thus, in V-Lab, the user cannot store an object for later use and therefore needs to move it in the lab, and since there is no inventory, the *pick up* action is unavailable.

### 3.4 Conceptual Modeling

The various simulated objects and instrument components in Onlabs and V-Lab are represented by MonoBehaviour *classes*; that is, classes with derive from a generic Unity class called MonoBehaviour which is attached on the GameObjects of the respective objects. For example, the *PhotonicMicroscope_LightIntensityKnob* is a class for the knob configuring the light intensity in any photonic microscope and is attached in any light intensity knob of microscope of that type. While in Onlabs the *PhotonicMicroscope_LightIntensityKnob* was a separate class, in V-Lab it derives from *LightIntensityKnob* class, a generic class for light intensity knobs of instruments of all kinds, which in turn derives from the *Knob* generic class. In other words, in V-Lab a broad hierarchy of classes has been designed, which requires a more abstract but also more efficient coding style.

Classes have *features* in the form of variables; for example, *PhotonicMicroscope_LightIntensityKnob* has a *Position* of type integer in the 1-24 range. They also have permitted *functions* on them; for example, the afore-mentioned class has a *turn* function, responsible for the turning of the knob. In V-Lab, the latter function is implemented in a generic form in *Knob* class, that is, the grandparent class of *PhotonicMicroscope_LightIntensityKnob*, and is inherited to it, while in Onlabs the respective function is implemented locally in the *PhotonicMicroscope_LightIntensityKnob* class. This generic way of function implementation in V-Lab is valid for the functions of other classes as well, with the exception of a few ones which are too specialized and need to be overridden.

## 4 EXTENDING ONLABS FOR XR2LEARN: V-LAB

As mentioned earlier, V-Lab is an extension of Onlabs developed for the purposes of the XR2Learn project that offers the same functionality as Onlabs in addition to advanced graphics, an improved user-interface and a modular architecture that enables a development and customization workflow for different scenarios. The architecture is illustrated in Figure 3.

The key concept underlying the architecture is the separation of the main simulation engine – the "V-Lab core" component – from scenario-specific assets. The core is available as a self-contained SDK consisting of a set of binary libraries for the Unity game engine, a client API, developer tools and documentation. Among the developer tools provided is a Unity editor tool that (currently) allows for exporting the contents of a scene to a file which can be loaded by the resulting executable at runtime. Thanks to that, V-Lab applications allow for a significant degree of flexibility as the scene setup can be modified and adapted to varying scenarios (in the same context and target area) after distribution without the need for work within the engine's development environment.

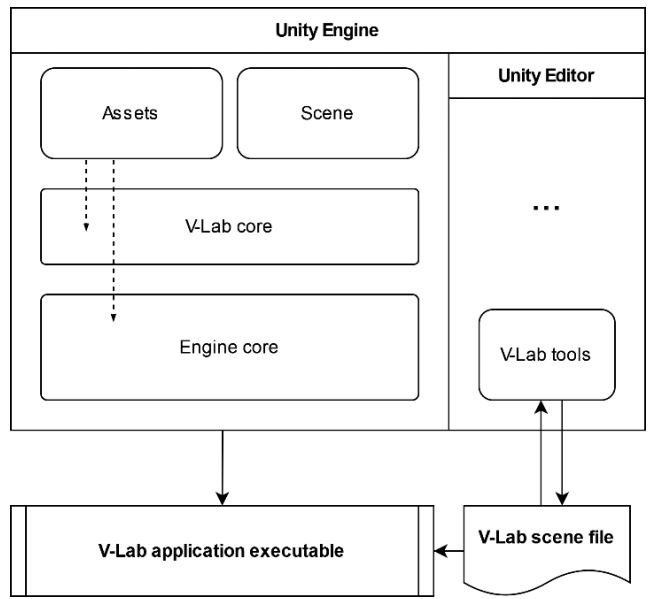

Figure 3. V-Lab architecture overview

The supported development workflow is as follows:

1. A new, empty project is created in the Unity game engine.
2. V-Lab core libraries are imported.
3. A generic or scenario-specific scene is designed or imported.
4. Scenario-specific assets are imported and incorporated into the scene or stored in the project as available elements for scene setup.
5. The overall scene setup, along with additional parameters, is exported as a V-Lab scene file via the editor tool.
6. Behaviour scripts are developed for the scenario-specific assets the V-Lab API and the engine's own APIs.
7. Executables are built for various target platforms and distributed along with V-Lab scene files.

All application-specific assets, including 3D models, audio and scripts, can be distributed as free and (if applicable) open-source assets, thus complying with the rationale and serving the purposes of XR2Learn. A beacon application based on V-Lab will indeed be developed and distributed as such according to the above-described workflow to illustrate the process of developing training scenarios based on V-Lab within the context of XR2Learn.

## 5 EVALUATION AND FUTURE WORK

Onlabs is constantly being evaluated by students, educators and other users and the evaluation bears encouraging results. Its most recent evaluation in 2021 verified, among others, that it increases the interest, the motivation and the engagement of the trainees, their confidence and certainty and their learning gain. It also helps them gain experimental skills by distance, while it consists of an effective solution of crisis situations like the recent Covid pandemic [4].

Future development and extension plans include revisiting the system's overall architecture towards a full-featured scenario development platform as well as a standalone scene editor tool that will enable visual scene manipulation by domain-experts at an application level.

**ACKNOWLEDGMENTS**

The work presented in this paper is part of the European project XR2Learn which has received funding by European Union's Horizon Europe Innovation Actions (Program Grant Agreement no. 101092851).